\documentclass[a4paper,12pt]{article}

\newcommand{\dfrac}[2]{\displaystyle{\frac{#1}{#2}}}

\def\qed{\rule{1mm}{3.8mm}}

\def\d  {\delta}
\def\la {\lambda}

\usepackage{theorem}

\theoremstyle{plain}
\newtheorem{Theo}{Theorem}
\newtheorem{Prop}{Proposition}
\newtheorem{Lemm}{Lemma}

\newcommand{\pl}{\partial}
\newcommand{\pder}[2]{\dfrac{\partial #1}{\partial #2}}
\newcommand{\oder}[2]{\dfrac{d #1}{d #2}}

\title{Nambu system associated with $n$-dimensional maps}
\author{Jun-ichi Yamamoto\footnote{E-mail: yjunichi@phys.metro-u.ac.jp}\\
\small{Department of Physics, Tokyo Metropolitan University} \\ 
\small{Minamiohsawa 1-1, Hachiohji, Tokyo 192-0397 Japan.}
}
\date{}
\begin{document}
\maketitle
\begin{center}
{\small\textbf{abstract}}\\
\vspace{3mm}

\begin{minipage}{.8\linewidth}
\small{
We studied that arbitrary 2-dimensional maps are Hamilton system if a initial value of map is a "time" variable. In this paper, we generalize this correspondence, and show that an $n$-dimensional map is a Nambu system in which one of initial values of the map play a role of "time" variable.
}
\end{minipage}
\end{center}

\section{Introduction}
In our previous paper \cite{SSYY}, we studied $2$-dimensional maps :
$$
(x^i_1,\;x^i_2)\;\longmapsto\;(x^{i+1}_1,\;x^{i+1}_2)
$$
and behaviors of point $(x^m_1(x^0_1,\;x^0_2),\;x^m_2(x^0_1,\;x^0_2))$ maped $m$ times repeatedly. The map is assumed to have its inverse and being differentiable. Changing point of view, let $t\equiv x^0_1$ be a independent "time" variable, $\la\equiv x^0_2$ be a fixed parameter, $X(t)\equiv x^m_1$ be a dependent coordinate variable and $P(t)\equiv x^m_2$ be also a dependent momentum variable. We denote by $J^{0,m}$ Jacobi matrix  of the map : $(x_1^0,x_2^0)\mapsto(x_1^m,x_2^m)$. In this view-point, we obtained the following result.
\begin{Theo}
Let $H$ be a function of $(X, P)$ given by
\begin{equation}
H(X,P)=\int^{\la}(\det J^{0m})d\la,
\label{Hamiltonian}
\end{equation}
and satisfying
$$
{\partial H\over\partial t}=0.
$$
Then the set of Hamilton's equations
\begin{equation}
{dX\over dt}={\partial H\over\partial P},\qquad 
{dP\over dt}=- {\partial H\over\partial X}
\label{Hamilton's equations}
\end{equation}
hold.
\end{Theo}
In order to support our claim we derived Hamiltonians corresponding to the H\'enon, KdV and $q$P$_{\mathrm{IV}}$ maps in \cite{SSYY}.
This view-point is based on studies of a discrete version of exact WKB analysis by Shudo and Ikeda. 

The aim of this paper is to generalize this mechanism, in order to find a dynamical system associated with $n$-dimensional map based on this view-point. In consequence, we will show that the corresponding dynamical system is a Nambu system.

Nambu system is a generalized Hamilton dynamical system which was introduced by Nambu, \cite{Nam,Ta}. This system is defined by $(n-1)$-Hamiltonians and Nambu brackets which reprace Poisson brackets in the ordinary Hamilton systems. Nambu brackets satisfy some properties such as skew-symmetry, Libnitz rule, fundamental identity and linear combination. Nambu system is useful tool, {\it ex.} deformation quantization \cite{DFST,DF}, dispersionless KP hierarchies and self-dual Einstein equation \cite{Gu}, {\it etc}. And one of the most famous problem is the Euler tops problem which have bi-Hamiltonian structure studied by Nambu \cite{Nam}. And more, commutators corresponding to Nambu bracket and algebra of it, called Nambu-Lie algebra, $n$-Lie algebra, $n$-ary Lie algebroid or Filippov algebroid are studied in recent \cite{DT,Fi,GM,GM1,GM2,ILMP,Vai,Vai2,Val}.

\section{$n$-dimensional maps}
Let us consider $n$-dimensional maps and inverse of them:
$$
s\;:\;(x_1^i,\dots,x_n^i)\;\longmapsto\;(x_1^{i+1},\dots,x_n^{i+1}),\qquad
s^{-1}\;:\;(x_1^{i+1},\dots,x_n^{i+1})\;\longmapsto\;(x_1^i,\dots,x_n^i),
$$
$$
x^{i+1}_j:=s(x_j^i)\equiv g_j(x^i_1,\dots,x^i_n),\qquad
x^i_j:=s^{-1}(x_j^{i+1})\equiv g_j^{-1}(x^{i+1}_1,\dots,x^{i+1}_n).
$$
where $g_j$'s are some differentiable functions.
We consider also Jacobi matrices associated wiht this maps :
$$
\begin{array}{lcccl}
J^{i,i+1}&:=&\left[\pder{x^{i+1}_j}{x^i_k}\right]&=&
\left[\begin{array}{ccc}
\pder{x^{i+1}_1}{x^i_1}&\cdots&\pder{x^{i+1}_1}{x^i_n}\\
\vdots & & \vdots\\
\pder{x^{i+1}_n}{x^i_1}&\cdots&\pder{x^{i+1}_n}{x^i_n}
\end{array}\right],
\vspace{3mm}\\
J^{i+1,i}&:=&\left[\pder{x^i_j}{x^{i+1}_k}\right]&=&
\left[\begin{array}{ccc}
\pder{x^i_1}{x^{i+1}_1}&\cdots&\pder{x^i_1}{x^{i+1}_n}\\
\vdots & & \vdots\\
\pder{x^i_n}{x^{i+1}_1}&\cdots&\pder{x^i_n}{x^{i+1}_n}
\end{array}\right].
\end{array}
$$
The Jacobi matrix $J^{0,m}$ is given by a product of them,
$$
J^{0,m}:=J^{0,1}\cdots J^{m-1,m},\qquad
J^{m,0}:=J^{m,m-1}\cdots J^{1,0},
$$
$$
J^{ij}J^{ji}=E,\qquad \textrm{($E$ : idntity matrix).}
$$
If we introduce notations $dx^i=(dx^i_1,\dots,dx^i_n)^T$ and $\pl^i=(\pl/\pl x^i_1,\dots,\pl/\pl x^i_n)^T$, the following results hold.
\begin{equation}
dx^{i+1}=J^{i,i+1}dx^i,\qquad
dx^i=J^{i+1,i}dx^{i+1},
\label{1-forms}
\end{equation}
\begin{equation}
\pl^{i+1}=(J^{i+1,i})^T\pl^i,\qquad
\pl^i=(J^{i,i+1})^T\pl^{i+1},
\label{1-vectors}
\end{equation}
where $T$ express a transposition.
\\

Here, let us change a point of view. We consider a $m$ times repeated map : $x^0\mapsto x^m$ where $x^i$ is a set of variables $(x_1^i,\dots,x_n^i)$. We also use notations as follows :
$$
(q_1,\dots,q_n)\equiv(x_1^m,\dots,x_n^m),\qquad
(\la_1,\dots,\la_{n-1},t)\equiv(x_1^0,\dots,x_n^0)
$$
$q_j(t)\;(j=1,\dots,n)$ are coordinates of an $n$-dimensional phase space, $\la_j\; (j=1,\dots,n-1)$ are fixed parameters and $t$ is a parameter which we consider as an independent "time" variable. In this view-point, the set of variables $q=(q_1,\dots,q_n)$ satisfy the following dynamical system.
\begin{Prop}\label{Map-Nambu}
Let $h=(h_1,\dots,h_{n-1})$ be a set of functions of $(q_1(t),\dots,q_n(t))$ given by
\begin{equation}
h_i=\int^{\la_i}(\det J^{0m})^{\frac{1}{n-1}}d\la_i,\qquad
i=1,\dots,n-1
\label{Hamiltonian}
\end{equation}
satisfying
\begin{equation}
\oder{h_i}{t}=0,\qquad
i=1,\dots,n-1.
\label{Conserving low}
\end{equation}
$$
$$
Then Nambu-Hamilton equations 
\begin{equation}
\oder{f}{t}=\{h_1,\dots,h_{n-1},f\}
\label{NH Equation}
\end{equation}
hold, where $f=f(q_1,\dots,q_n,t)$ is a certain function. 
\end{Prop}
If $f=q_i$ then $(\ref{NH Equation})$ is an equation of motion. In $(\ref{NH Equation})$, Nambu brackets is defined by
\begin{equation}
\{f_1,\dots,f_n\}=\pder{(f_1,\dots,f_n)}{(q_1,\dots,q_n)}.
\end{equation}
Here we assume the existence of the inverse map $s^{-1}$, such that $h_j$'s are considered as functions of $q_j$'s through $\la=\la(q)=(s^{-1})^m(q)$.

For simplicity, we define some symbols before proof. 
$H$ is a Jacobi matrix of $(h_1,\dots,h_n)$ given by
$$
H_q:=\left[\pder{h_j}{q_k}\right],\qquad
H_\la:=\left[\pder{h_j}{\la_k}\right], \qquad j,k=1,\dots,n,
$$
and $\tilde{H}_q$ is a cofactor matrix of $H_q$. Namely the $(j,k)$-element of $\tilde{H}_q$ is the $(j,k)$-cofactor of $H_q$. Here, we set formally $\la_n=t$ and 
$$
h_n:=\int^{\la_n}(\det J^{0m})^{\frac{1}{n-1}}d\la_n
$$
\\
Then these matrices satisfy the following Lemma.
\begin{Lemm}\label{3rel}
Let us consider the Nambu-Hamilton equation given by
\begin{equation}
\pder{f}{\la_k}=\{h_1,\dots,h_{k-1},f,h_{k+1},\dots,h_n\},\qquad k=1,\dots,n,
\label{GNH Equation}
\end{equation}
where $\la_j\;(1\le j \le n, j\neq k)$ are fixed parameters, 
$\la_k$ is a independent parameter, 
$q_j\;(1\le j \le n)$ are depnendent parameters $q_j(\la_k)$, 
$h_j\;(1\le j \le n)$ are hamiltonians without $h_k$ and 
$f,f_j\;(1\le j \le n)$ are arbitrary functions $f_j(q_1,\dots,q_n,\la_1,\dots,\la_n)$.

Then, for above Jacobi matrices $H_q$, $H_\la$, $J^{0m}$, the cofactor matrices $\bar H_q$ and $\bar H_\la$, 
following three relations hold.
\begin{enumerate}
	\item $H_\la= H_q J^{0,m}$,
	\item $J^{0,m}=\tilde H_q^T$,
	\item $H_q=(\det\tilde H_q)^{\frac{1}{n-1}}(\tilde H_q^T)^{-1}$.
\end{enumerate}
\end{Lemm}

\noindent
\textbf{Proof of Lemma \ref{3rel}} : 

\noindent
1. $H_\la= H_q J^{0,m}$. Using (\ref{1-vectors}), 
$$
\left[\begin{array}{c}\pl h_j/\pl \la_1 \\ \vdots \\ \pl h_j/\pl \la_n\end{array}\right]
=(J^{0,m})^T
\left[\begin{array}{c}\pl h_j/\pl q_1 \\ \vdots \\ \pl h_j/\pl q_n\end{array}\right]
$$
Hence, 
$$
H_\la^T=(J^{0,m})^TH_q^T.
$$
Transposing this, therefore, the relation $H_\la= H_q J^{0,m}$ hold.\hfill \qed \\

\noindent
2. $J^{0,m}=\tilde H_q^T$. Substituting $q_j$ to $f$ in Nambu-Hamilton equation (\ref{GNH Equation}),
$$
\pder{q_j}{\la_k}=\{h_1,\dots,h_{k-1},q_j,h_{k+1},\dots,h_n\}
$$
Then r.h.s. of this is a $(k,j)$ cofactor, because 
$$
\pder{(h_1,\dots,h_{k-1},q_j,h_{k+1},\dots,h_n)}{(q_1,\dots,,q_n)}=
(-1)^{k+j}\pder{(h_1,\dots,h_{k-1},h_{k+1},\dots,h_n)}{(q_1,\dots,q_{j-1},q_{j+1},\dots,q_n)}=
\tilde{h}_{k,j}.
$$
And l.h.s. is one of entries of $J^{0,m}$. 
Hence we obtain $J^{0,m}=\tilde H_q^T$.\hfill \qed \\

\noindent
3. $H_q=(\det\tilde H_q)^{\frac{1}{n-1}}(\tilde H_q^T)^{-1}$. 
It is well known that an arbitrary $n\times n$ matrix $A$ and its cofactor matrix $\tilde A$ satisfy the following relation. 
$$
A\tilde A^T=\tilde A A^T=(\det A)E.
$$
Since this relation derives
$$
\det\tilde A = (\det A)^{n-1},
$$
the matrix $A$ can be expressed by $\tilde A$ as follows :
$$
A=(\det\tilde A)^{\frac{1}{n-1}}(\tilde A^T)^{-1}.
$$
If $A=H_q$, the relation 3 holds.\hfill \qed \\

\noindent
\textbf{Proof of Proposition \ref{Map-Nambu}} : 
We assume that maps $s$, $s^{-1}$ and their explicit forms $g_j$ are given.
\\ \\
($\mathrm{i}$) : 
We must show that functions $h_j$ satisfy Nambu-Hamilton equation $(\ref{NH Equation})$ if $h_j$ are given by $(\ref{Hamiltonian})$, because $h_j$'s are given by explicit functions $g_j$. Hence, we must check a compatibility between $(\ref{Conserving low})$ and $(\ref{NH Equation})$. Substituting $h_j$ to $(\ref{NH Equation})$, 
$$
\oder{h_j}{t}=\{h_1,\dots,h_{n-1},h_j\}=0 \quad \textrm{ if }\quad j\neq n
$$
because Nambu bracket is a Jacobian. Therefore if the functions $h_j$, called Hamiltonians, are given by maps $s$ and $s^{-1}$ with $(\ref{Hamiltonian})$, then $h_j$'s satisfy Nambu-Hamilton equation.
\\ \\
($\mathrm{ii}$) : 
We will show that if Nambu-Hamilton equation is given by $(\ref{NH Equation})$ and maps $s$, $s^{-1}$ are given, then functions $h_j$ are given by $(\ref{Hamiltonian})$. 
Using the three relations of Lemma \ref{3rel},
$$
H_\la=H_qJ^{0,m}=(\det\tilde H_q)^{\frac{1}{n-1}}(\tilde H_q^T)^{-1}\tilde H_q^T=(\det J^{0,m})^{\frac{1}{n-1}}E.
$$
Since r.h.s. is a diagonal matrix, we obtain $h_j$ as follow :
$$
\pl_{\la_k}h_j=(\det J^{0,m})^{\frac{1}{n-1}}\d_{j,k}\quad \Longrightarrow\quad 
h_j=\int^{\la_j}(\det J^{0,m})^{\frac{1}{n-1}}d\la_j
$$
Therefore, if maps and Nambu system are given then Hamiltonian $h_j$ are given by $(\ref{Hamiltonian})$. 
\\ \\
($\mathrm{iii}$) : 
If maps $s$ and $s^{-1}$ has been given, then there exist Nambu system corresponding to maps because of ($\mathrm{i}$) and ($\mathrm{ii}$). The Nambu system have Nambu-Hamilton equation $(\ref{NH Equation})$ and Hamiltonians $(\ref{Hamiltonian})$. \hfill \qed \\

The generalized Nambu-Hamilton equation $(\ref{GNH Equation})$ is not dynamical equation. If we select one independent variable $\la_k$ as a time variable, then Nambu-Hamilton dynamical equation is given by
$$
\oder{f}{\la_k}=\{h_1,\dots,h_{k-1},f,h_{k+1},\dots,h_n\}.
$$
This equation is also Nambu-Hamilton equation, and Hamiltonians are $h_j,\;(j\neq k)$ but $h_k$ is not Hamiltonian. We can choose one independent variable in parameters $(\la_1,\dots,\la_n)$ on Nambu system, freely.

On the Nambu system, explicit functions $g_j$ of maps are solutions of Nambu dynamics, because this functions 
$$
q_j(t)=g_j^m(\la_1,\dots,\la_{k-1},t,\la_{k+1},\dots,\la_n)
$$
are depend on $(n-1)$-constants and one independent variable, where $g^m_j$ is a explicit form of $m$ time repeated maps of $s$.

In the special case of $(\det J^{0m})=1$, $(n-1)$-constants are $(n-1)$-Hamiltonians, since
$$
h_j=\int^{\la_j}d\la_j=\la_j.
$$
And the map $s$ is a canonical transformation or a $n$-dimensional volume preserving transformation, since
$$
dx_1^{i+1}\wedge\cdots\wedge dx_n^{i+1}=dx_1^i\wedge\cdots\wedge dx_n^i,
$$
and
$$
dq_1\wedge\cdots\wedge dq_n = d\la_1\wedge\cdots\wedge dt \wedge\cdots\wedge d\la_n= dh_1\wedge\cdots\wedge dt \wedge\cdots\wedge dh_n.
$$
So, $(h_j,t)$ is a set of canonical conjugate variables.

In our sense, a independent "time" value is a initial value. This mean that the response of changes of a initial value in discrete systems can be investigated with Nambu mechanics in continuum systems because of this Nambu-map correspondence. 

\section{Example}
\subsection{Lotka-Volterra map}
Discrete Lotka-Volterra equation : 
$$
\bar{x}_k\left(1+\bar{x}_{k-1}\right)=x_k(1+x_{k+1}),\qquad k=1,2,3
$$
have a $3$-dimensional map and its inverse
$$
\bar{x}_k=x_k\dfrac{1+x_{k+1}+x_{k+1}x_{k+2}}{1+x_{k+2}+x_{k+2}x_k}
,\qquad
x_k=\bar x_k\dfrac{1+\bar x_{k+2}+\bar x_{k+2}\bar x_{k+1}}{1+\bar x_{k+1}+\bar x_{k+1}\bar x_k}
$$
under periodic boundary condition $x_{k+3}=x_k$, where $\bar x_k=x_k^{i+1}$, $x_k=x_k^i$. Jacobi matrix $J^{i,i+1}$ is given by
$$
\left[\begin{array}{ccc}
\dfrac{(1+x_2+x_2x_3)(1+x_3)}{(1+x_3+x_3x_1)^2}&
-\dfrac{x_2(1+x_2+x_2x_3)}{(1+x_1+x_1x_2)^2}&
\dfrac{x_3(1+x_2)}{1+x_2+x_2x_3}\vspace{3mm}\\
\dfrac{x_1(1+x_3)}{1+x_3+x_3x_1}&
\dfrac{(1+x_3+x_3x_1)(1+x_1)}{(1+x_1+x_1x_2)^2}&
-\dfrac{x_3(1+x_3+x_3x_1)}{(1+x_2+x_2x_3)^2}\vspace{3mm}\\
-\dfrac{x_1(1+x_1+x_1x_2)}{(1+x_3+x_3x_1)^2}&
\dfrac{x_2(1+x_1)}{1+x_1+x_1x_2}&
\dfrac{(1+x_1+x_1x_2)(1+x_2)}{(1+x_2+x_2x_3)^2}
\end{array}
\right]
$$
and its Jacobian and inverse are the following
$$
\det J^{i,i+1} = 1,\qquad \det J^{i+1,i}=1
$$
because $J^{i,i+1}J^{i+1,i}=E$. Now, we will consider the simplest case $m=1$. Setting up variables as follows, 
$$
(h_1,h_2,t)=(\la_1,\la_2,\la_3)=(x_1^0,x_2^0,x_3^0),\qquad
(q_1,q_2,q_3)=(x_1^1,x_2^1,x_3^1),
$$
satisfy the following Nambu system.
\begin{itemize}
	\item Equations of motion
	$$
	\oder{q_1}{t}=\pder{(h_1,h_2)}{(q_2,q_3)},\qquad
	\oder{q_2}{t}=-\pder{(h_1,h_2)}{(q_1,q_3)},\qquad
	\oder{q_3}{t}=\pder{(h_1,h_2)}{(q_1,q_2)},
	$$
	\item Hamiltonians
	$$
	h_1=q_1\dfrac{1+q_3+q_3q_2}{1+q_2+q_2q_1},\qquad
	h_2=q_2\dfrac{1+q_1+q_1q_3}{1+q_3+q_3q_2},
	$$
	\item Solutions
	$$
	q_1(t)=h_1\dfrac{1+h_2+h_2t}{1+t+h_1t},\qquad
	q_2(t)=h_2\dfrac{1+t+h_1t}{1+h_1+h_1t},\qquad
	q_3(t)=t\dfrac{1+h_1+h_1h_2}{1+h_2+h_2t},
	$$
	\item Explicit forms of equations of motion 
	$$
	\begin{array}{l}
	\oder{q_1}{t}
	=\frac{-q_1(1+q_1+q_1q_3)}{(1+q_2+q_2q_1)(1+q_3+q_3q_2)},
	\vspace{3mm}\\
	\oder{q_2}{t}
	=\frac{q_2(1+q_2)(1+q_1+q_1q_3)}{(1+q_2+q_2q_1)(1+q_3+q_3q_2)},
	\vspace{3mm}\\
	\oder{q_3}{t}
	=\frac{(1+q_3)(1+q_1+q_1q_3)}{(1+q_2+q_2q_1)(1+q_3+q_3q_2)}.
	\end{array}
	$$
\end{itemize}

\end{document}